\documentclass[useAMS]{mn2e}
\usepackage{psfig}
\usepackage{graphicx}

\title{Multiband optical monitoring of the blazars S5 0716+714 and 
BL Lacertae }

\author[C. S. Stalin et al. ] 
{C.S.\ Stalin$^{1,2,3}$, Gopal-Krishna$^{3,4}$, Ram Sagar$^{2}$, Paul J.\ Wiita$^{5}$, V.\ Mohan$^{1}$, A.K.\ Pandey$^{2}$  \\  
$^{1}$ Inter-University Centre for Astronomy and Astrophysics (IUCAA), Post Bag No. 4, Ganeshkhind, Pune 411 007, India  \\
$^{2}$ Aryabhatta Research Institute of Observational Sciences (ARIES), Manora Peak, Nainital 263 129, India \\
$^{3}$ Institut d'Astrophysique de Paris, 98bis, bd Arago, 75014, Paris, France \\
$^{4}$ National Centre for Radio Astrophysics, TIFR, Pune University Campus, Post Bag 3, Pune 411
007, India\\
$^{5}$ Department of Physics \& Astronomy, Georgia State University, P.O. Box 4106, Atlanta,
Georgia 30302-4106, USA}

\date{Accepted xxx xxx 2005.
      Received xxx xxx 2005;
      in original form xxx July 2005}

\pagerange{\pageref{firstpage}--\pageref{lastpage}}
\pubyear{xxx}

\begin{document}

\maketitle

\label{firstpage}

\begin{abstract}
 We report results of multiband optical photometric monitoring of two 
 well known blazars, S5 0716+714 and BL Lacertae, carried out during 
  1996 and 2000$-$01 with an aim to study optical 
 variations on time scales ranging from minutes to hours and
 longer. The light curves were derived relative to comparison 
 stars present on the CCD frames. Night to night intensity variations of 
 $\ge$ 0.1 mag were observed in S5 0716+714 during a campaign of about 
 2 weeks in 1996. A good correlation between the lightcurves in different 
optical bands was 
found for both inter-night and intra-night observations.
 In all, two prominent events of intra-night optical 
 variability (INOV) were detected in S5 0716+714. Each of these rapidly
 varying segments
 of the lightcurves can be well fitted with an exponential intensity
 profile whose rate of variation is essentially the same in both cases. 
 Our long-term monitoring data of S5 0716+714 showed a distinct flare 
 around JD 2451875 which can be identified in the {\it BVRI} bands. 
This flare coincides with the brightest phase recorded during
1994$-$2001 in the long-term lightcurves reported by Raiteri et al.\ (2003).
No evidence for the object to become bluer when brighter was noticed
 on either inter-night or intra-night time scales.
On the other hand, our essentially simultaneous multiband optical observations
of BL Lacertae in October 2001 showed flux variations that were not 
achromatic. 
This
blazar definitely was found to become bluer when brighter on 
intra-night time scales and there is a less significant trend of the same type
on inter-night time scales. 
Based on five nights of observations during a week, BL Lacertae showed
a peak night-to-night variability of $\sim$ 0.6 mag in {\it B}.
Thus, we found that the present optical observations of the two 
prominent blazars, made with similarly high sensitivity, reveal a contrasting behaviour in terms of the dependence of 
spectral hardening with increasing brightness, at least on intra-night, and possibly also on
inter-night, time scales.
\end{abstract}

\begin{keywords}
BL Lacertae objects: individual: S5 0716+714, BL Lacertae -- galaxies: active
-- galaxies: photometry
\end{keywords}

\section{Introduction}

Active galactic nuclei (AGNs) are known to be variable on different
time scales across the electromagnetic spectrum. Photometric
studies of intensity variations in AGNs provide a uniquely
powerful tool for investigating the processes occurring in the
vicinity of their central engine.
In particular, studies of very rapid intensity variations, or intra-night optical
variability (INOV), where the 
variations have amplitudes of a few hundredths of a magnitude 
on hour-like or shorter time scales, 
enable us to probe their
innermost nuclear cores, on the scales of microarcseconds. 
Blazars define a class of AGNs  made up of optically violent variable quasars,
high polarization quasars and BL Lac objects. They show high polarization
and violent variability at optical wavelengths, and in the radio band contain
 compact flat spectrum sources
which often exhibit apparent superluminal motion and a
high polarization level. These characteristics are generally attributed to
synchrotron emission from a relativistic jet with the jet axis oriented at
small angles to the observer's line of sight.

Theoretical explanations for the origin of
INOV  can be broadly divided into intrinsic and extrinsic
categories. One such extrinsic mechanism is refractive interstellar scintillation 
(e.g., Kedziora-Chudczer et al.\ 1997), 
though this is relevant only in the radio band. 
Another proposed extrinsic mechanism is superluminal microlensing 
(Gopal-Krishna \& Subramanian 1991), which can explain the rapidity of variations,
though it is unlikely to apply to a large fraction of blazars.
The dominant basic model for intrinsic variability invokes shocks propagating through the
jet (e.g., Blandford \& K\"{o}nigl 1979; Marscher \& Gear 1985).  Models designed to
explain INOV based
on jets involve turbulence behind the jet (Marscher, Gear \& Travis 1992), light echo
effects (Qian et al. 1991), helical filaments (Rosen 1990) or changing
directions of the shocks in the jet (Gopal-Krishna \& Wiita 1992; Nesic et al.\ 2005). 
The effectiveness of
all these models are enhanced by relativistic effects in the jet, 
especially when the viewing angle is small. Another family of intrinsic
explanations invokes numerous flares or hot spots on the surface of (or in the corona
above) the accretion disk (e.g., Mangalam \& Wiita 1993). Similar models have  
been proposed to explain the X-ray variations of AGN 
(e.g., Abramowicz et al.\ 1992).  
While such disk perturbations are unlikely to
dominate blazar variability they may propagate into or otherwise affect
jet dominated emission (e.g., Wiita 2005).

\subsection{S5 0716$+$714}
S5 0716$+$714, classified as a BL Lac object, was discovered in the Bonn-NRAO
Radio Survey (K\"{u}hr et al.\ 1981) of flat spectrum radio sources with
a 5 GHz flux greater than 1 Jy. 
It is believed to be at a redshift z $>$ 0.3 (Wagner et al. 1996) due
to the lack of the detection of its host galaxy; however, a lower
value of z $\sim$ 0.1 has also been suggested recently (Kadler et al. 2004).
This source, an intra-day variable
at radio and optical wavelengths as well as a $\gamma$-ray emitter
(Wagner et al.\ 1996), is a favourite target for variability studies. In a
4 week long monitoring campaign in February 1990, this blazar showed an
abrupt transition in its variability pattern from a higher flux level 
with roughly
1-day time scale during the first week, to a lower flux level with
 approximately 7-day time scale for the remainder of the campaign,
both at optical and centimeter wavelengths (Quirrenbach et al.\ 1991).
Another interesting result is that the radio spectral index
of this blazar was found to correlate with intra-night optical
variations such that a flattening of the radio spectrum near the optical
maxima was found for several quasi-cycles (Qian et al.\ 1996; Wagner et al.\ 1996). 
The intriguing evidence for
correlated variability in the radio and optical 
ranges (Quirrenbach et al. 1991; Wagner et al. 1996) certainly points to an
intrinsic origin of the variations in this blazar.
These
authors further inferred that the {\it R} band light curves evinced a significant
signature of flickering on time scales as short as 15 minutes,  much
faster than the apparent quasi-periodic time scales.

Raiteri et al. (2003) present results on multi-band optical observations ranging
from 1994$-$2001 and also present 
radio observations spanning more than 20 years. They report that the
long term optical brightness variations of this source appears to have a 
characteristic timescale
of $\sim$3.3 yrs.  Ghisellini et al.\ (1997) and Sagar et al.\ (1999) too have reported
results of monitoring of this blazar in BVRI colours during
February to April 1994. Their campaigns recorded a few events of
INOV with an amplitude of about 5\% within
a few hours, superposed on slower variations.

In order to further investigate the INOV behaviour of this blazar,
we carried out a two
weeks long optical monitoring campaign during April 1996. These
observations covered a total baseline of 16 days and provided a fairly
dense temporal coverage on most of the nights. These data, in conjunction
with our additional observations during 2000$-$01, are used here to 
investigate both short-term and long-term variability of this prominent blazar.

\subsection{BL Lacertae}
BL Lac (2200$+$420), the prototype of
this class of AGNs, was identified by Schmitt (1968) 
with the radio source VRO 42.22.01, and its spectrum was found to be
featureless by Viswanathan (1969). However, a broad $H{\alpha}$ line with
EW $>$ 5 \AA ~was later detected by Vermeulen et al.\ (1995). It is found to
be embedded in an elliptical host galaxy (Wurtz, Stocke \& Yee 1996) with $z = 0.069$.
Though large amplitude variations
on time scales ranging from days to decades (Webb et al.\ 1988) and rapid variations
of 0.1 mag/hr (Racine 1970) were already known, the advent of CCDs used as N-star
photometers conclusively
demonstrated INOV in BL Lac (Miller et al.\ 1989). These
authors report rapid changes in the {\it V} band flux on
time scales as short as 1.5 hrs. 
It brightens almost every year and was reported
to be in the active state in 2001 by Mattei et al.\ (2001). 
We carried out intra-night BVR monitoring of this blazar 
for 5 nights in October 2001. 

\subsection{Spectral variability in blazars}
Although optical variability on hour-like time scales is now a 
well established phenomenon for blazars, its relationship to
the long-term variability remains unclear. Possible clues
could come from monitoring the optical spectrum for correlations 
with brightness. S5 0716+714 has been intensively studied for variability 
across the electromagnetic spectrum on various 
time scales (see Wu et al. 2005 and references therein; Nesci et al. 2005).
On inter-night time scales a bluer when brighter correlation was found when 
the object was in an active or flaring state, but this trend was absent 
during its relatively quiescent state (Wu et al. 2005). Also, during one
night's monitoring, Wu et al. (2005) noticed
no spectral hardening with brightness. 

During a 11-day long intra-night monitoring campaign of BL Lac during 
a major outburst in July 1997, clear evidence for the object
to become bluer when brighter was noted by Clements \& Carini (2001).
However, they interpreted this as an artefact of an increased
contribution from the host galaxy when the blazar was fainter.
In contrast, based on their BRI photometry of BL Lacertae on 
5 nights during 1999 and 2001, Papadakis et al.\ (2003) 
argued that spectral hardening with increasing brightness is evident
even after the contribution from the host galaxy is subtracted, hinting 
that the effect is intrinsic to the blazar. The same
significant trend was independently reported by Villata et al.\ (2004) 
from the intra-night monitoring of BL Lacertae in UBVRI
bands. They also noted that this colour--brightness 
correlation is much weaker for the long-term brightness 
variations, which were nearly achromatic on a few day-like time scale.
The results of long-term BVRI monitoring between 1995 and 1999 of
8 BL Lac objects, including BL Lacertae and S5 0716$+$714, by D'Amicis et al.\ (2002)
were further analysed by Vagnetti et al.\ (2003) who found a tendency for bluer
colour at higher luminosity for all of them; however, for their
dataset, BL Lac showed a fairly strong correlation while 0716$+$714 showed
a weaker one.

\begin{figure*}
\psfig{file=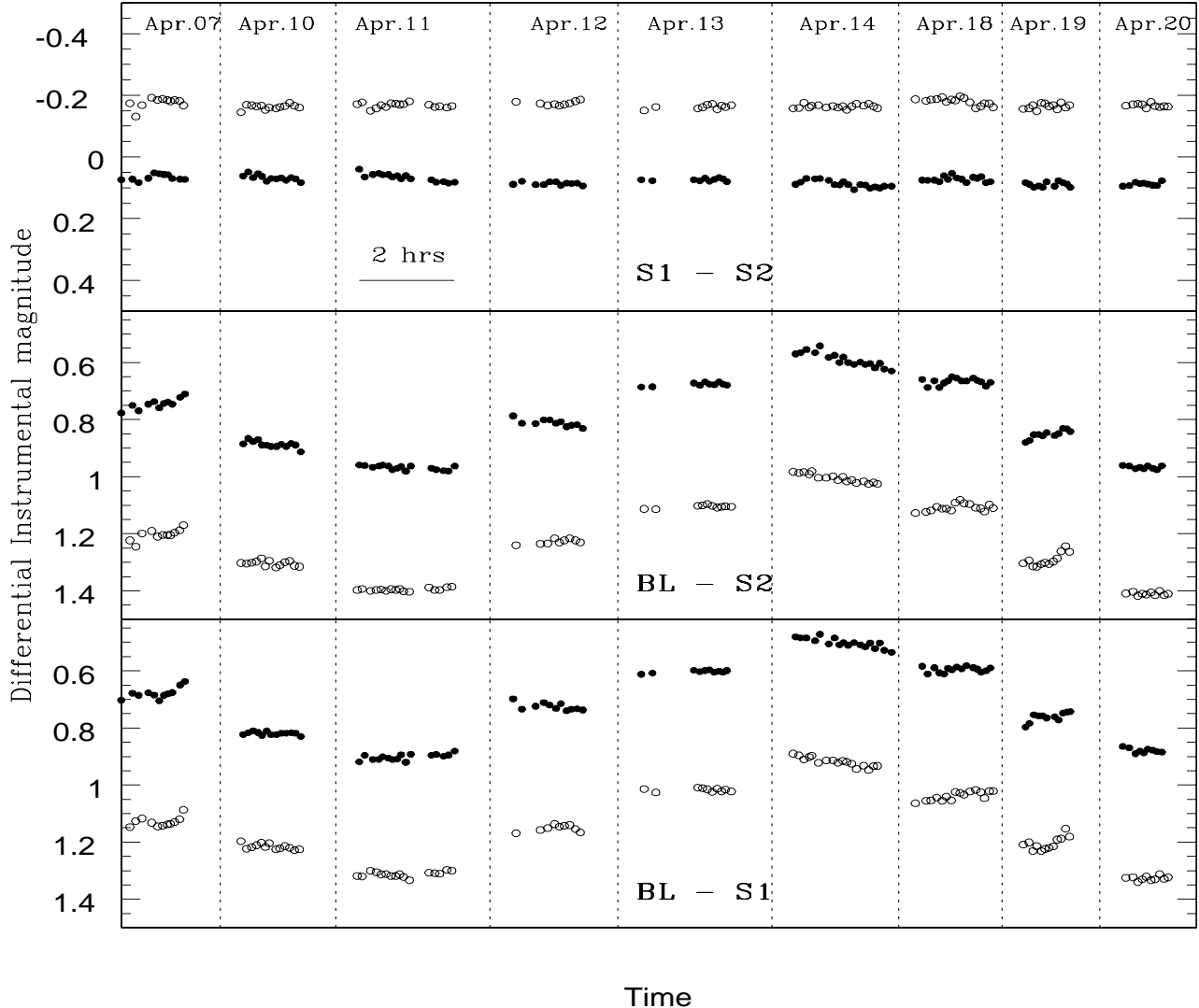,width=19cm,height=16cm}  
\caption{DLCs for S5 0716+714 and the two comparison stars (Sect. 3).
Filled and open circles denote the differential magnitude in {\it R} and V
bands, respectively. The {\it V} band DLCs have been shifted by 
$+$0.4 mag (between blazar and comparison stars) and $-$0.25 mag (between
comparison stars themselves), respectively. The
time scale for all DLCs is indicated by the horizontal bar, 
with actual times given in Table 1.}
\end{figure*}

If the trend of spectral hardening with increasing brightness is confirmed to 
be (nearly) universal, at least for the variations on hour-like time scales, one
interesting explanation posits short-term fluctuations of only the electron injection 
spectral index (e.g., Mastichiadis \& Kirk 2002; B\"{o}ttcher \& Reimer 2004). 
However, some blazars are found to show anomalous spectral 
behaviour (see Ram\'{i}rez et al.\ 2004 and references therein). For example, 
PKS 0736$+$017 showed a tendency for its spectrum to become redder when 
brighter (Ram\'{i}rez et al.\ 2004), both on inter-night and intra-night 
time scales. The multiband optical observations of the prominent 
blazars S5 0716+714 and BL Lacertae reported here were carried out to shed 
more light on the possible dependence of optical spectrum 
on the brightness state of blazars. 
In the following section we describe the observations and data reduction. 
Sect.\ 3 presents the analysis of the lightcurves, and the results are 
summarized in Sect.\ 4.

\hspace*{-4.0cm}
\section{Observations and Reductions}
The observations were carried out at the f/13 Cassegrian focus of the
104 cm reflector of the Aryabhatta Research Institute of observational 
sciencES (ARIES), Nainital, India. Observations during
1996 were performed using a 1024 $\times$ 1024 pixels$^2$ CCD whereas the
2000$-$01 observations employed a 2048 $\times$ 2048 pixels$^2$ chip.
The pixel size of 24 $\times$ 24 $\mu m^2$ of both the CCD systems
corresponds to 0.37 $\times$ 0.37 (arcsec)$^2$ on the sky. Observations were done in
2 $\times$ 2 binned mode to improve S/N. Exposure times varied between 60 
and 600 sec, depending on the brightness of the object and sky conditions.
Several bias frames were taken intermittently during observations and
twilight sky flats were taken to correct for pixel to pixel
variations on the chip.

Initial processing of the images, including bias subtraction and flat
fielding, was done using IRAF\footnote{IRAF is distributed by the National
Optical Astronomy Observatories, which is operated by the Association of 
Universities for Research in Astronomy Inc.\ under contract to the National
Science Foundation.} whereas cosmic ray removal was
carried out using MIDAS\footnote{Munich Image Data Analysis System; trademark
of the European Southern Observatory (ESO)}. Photometry of the flat fielded frames 
were carried
out using DAOPHOT (Stetson 1987). Typical seeing during the observations
was around 2.0$^{\prime\prime}$. The log of observations for
INOV is given in Table 1.  In order to detect weak 
fluctuations we have performed differential photometry, generally ensuring
that the locations of the blazar and the comparison stars on the CCD frame 
do not change by
more than a few pixels from exposure to exposure during a night.
In all, S5 0716$+$714 was monitored on 11 nights between 4 to
20 April 1996 and BL Lacertae was observed on 5 nights between 
19 and 25 October 2001.

\begin{table}
\caption{Log of intranight  observations}
\begin{tabular}{llll} \hline
\noalign{\smallskip}
Date  &  Object   &  Duration  &  Filter(s)         \\
      &           &   UT (hrs)    &                  \\
\noalign{\smallskip} \hline
\noalign{\smallskip}
04 Apr. 96  &  S5 0716$+$716  &  15.5$-$17.1  &  V    \\
05 Apr. 96  &  S5 0716$+$714  &  16.1$-$17.1  &  R    \\
07 Apr. 96  &  S5 0716$+$714  &  15.0$-$16.4  &  V,R  \\
10 Apr. 96  &  S5 0716$+$714  &  15.3$-$16.5  &  V,R  \\
11 Apr. 96  &  S5 0716$+$714  &  14.0$-$16.0  &  V,R  \\
12 Apr. 96  &  S5 0716$+$714  &  14.9$-$16.4  &  V,R  \\
13 Apr. 96  &  S5 0716$+$714  &  14.1$-$15.9  &  V,R  \\
14 Apr. 96  &  S5 0716$+$714  &  14.8$-$16.8  &  V,R  \\
18 Apr. 96  &  S5 0716$+$714  &  14.7$-$16.3  &  V,R  \\
19 Apr. 96  &  S5 0716$+$714  &  15.7$-$16.7  &  V,R  \\
20 Apr. 96  &  S5 0716$+$714  &  14.7$-$15.5  &  V,R  \\
19 Oct. 01  &  BL Lacertae  &  18.4$-$20.4  &  B,V,R \\
20 Oct. 01  &  BL Lacertae  &  18.8$-$20.5  &  B,V,R \\
21 Oct. 01  &  BL Lacertae  &  18.8$-$20.7  &  B,V,R \\
22 Oct. 01  &  BL Lacertae  &  18.4$-$20.4  &  B,V,R \\
25 Oct. 01  &  BL Lacertae  &  19.8$-$21.1  &  B,V,R \\
\noalign{\smallskip}
\hline
\end{tabular}
\end{table}

\begin{figure}
\psfig{file=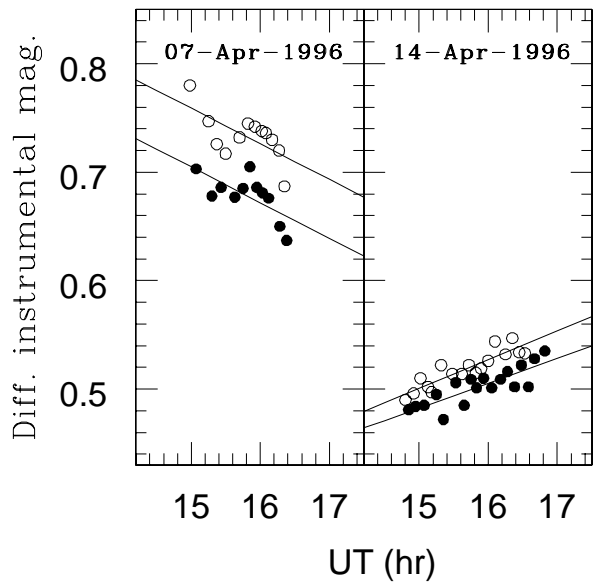,width=9cm,height=9cm}  
\caption{Linear unweighted regression fits to the differential magnitudes
of S5 0716$+$714 for the two nights when the object showed INOV. Filled and
open circles denote {\it R} and {\it V} bands, respectively.}
\end{figure}

\section{Light curve analysis}
The instrumental magnitudes resulting from DAOPHOT were used to construct the
differential light curves (DLCs). These represent the 
instrumental magnitudes of the blazar relative to two comparison
stars on the CCD frame. The standard deviation ($\sigma$) of the DLC for 
the comparison stars gives
a measurement of the observational errors 
(e.g., Jang \& Miller 1995). The comparison stars used for the
differential photometry of S5 0716$+$714 are the stars 5 and 6 identified
by Villata et al.\ (1998). 
For BL Lacertae,
stars B and C of Smith et al.\ (1985) are used for the differential photometry. 
The comparison stars are chosen so as to minimize differences between their
instrumental brightnesses and colours and those of 
the corresponding blazars.

\begin{figure}
\hspace*{-0.7cm}\psfig{file=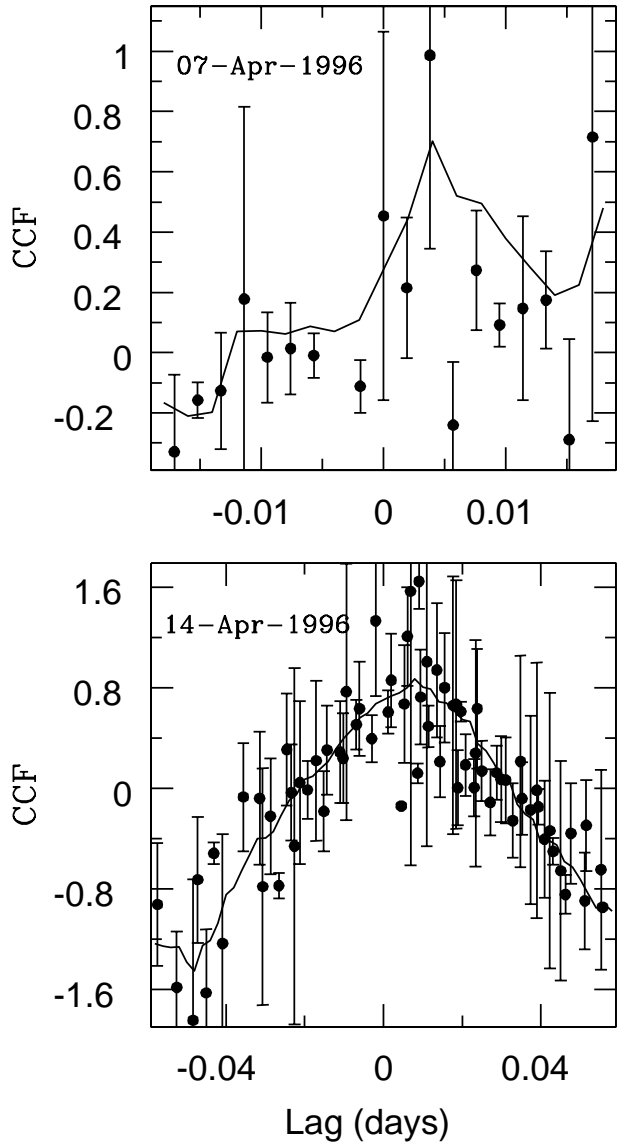} 
\caption{Cross correlation function for the {\it R} and {\it V} band lightcurves of S5 0716$+$714 
taken on 07 April 1996 (top panel) and 14 April 1996 (bottom panel). 
Filled circles with error bars refer to the DCF, whereas the solid line
is calculated from the ICCF (see Sect. 3.1.1).}
\end{figure}

In claiming that a source is variable
we employed the frequently used criterion adopted by  Jang and Miller (1997). The 
 confidence level of variability, when observed, is defined 
 as C = $\sigma_T/\sigma$ where
$\sigma_T$ is the standard deviation of the blazar DLC. The adopted variability
criterion requires that C $\ge$ 2.576 for the object to be classified as 
variable at 99\% confidence level (Jang \& Miller 1997; Stalin et al.\ 2004). 
On a night when three or more filters are employed the
 source is called variable only if C $\ge$ 2.576 in at least 
two filters.

\begin{figure}
\hspace*{-0.4cm}\psfig{file=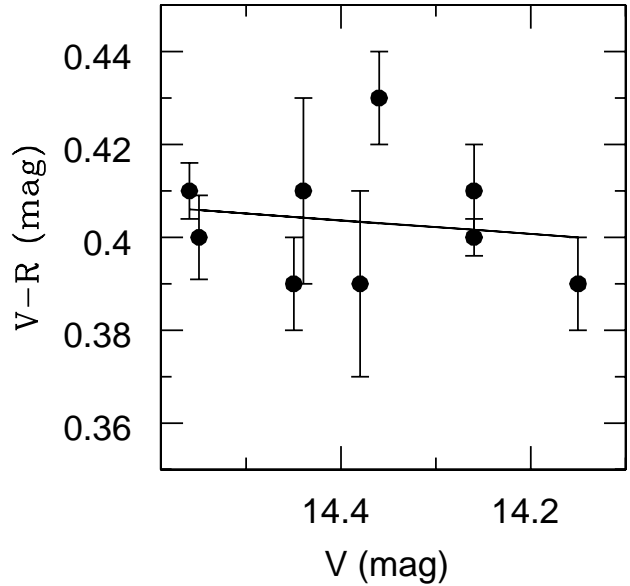} 
\caption{Inter-night colour magnitude diagram (CMD) for ~S5 0716$+$714. Th solid 
line is the linear least-squares fit to the data.}
\end{figure}

\begin{figure}
\hspace*{-0.2cm}\psfig{file=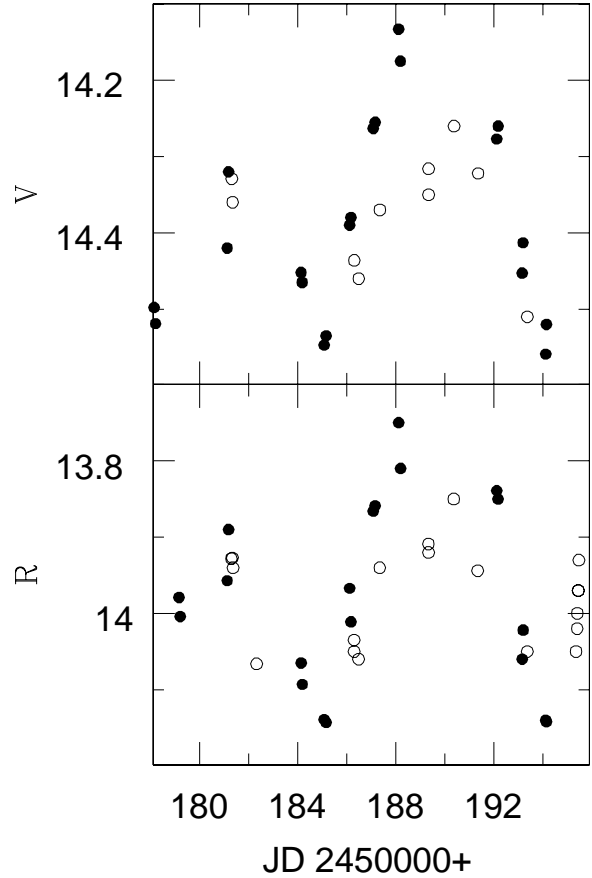} 
\caption{Inter-night variability of S5 0716$+$714 during our observations
(filled circles). For
each night two data points of differential magnitude are 
plotted (see Sect.\ 3.1.3). Note that the first pair of data points in 
the upper and lower panels refer to two consecutive nights (Table 1) and
these are not shown in Fig.\ 1. The open circles are taken from
the long-term monitoring campaign by Raiteri et al.\ (2003).}
\end{figure}

\begin{figure}
\hspace*{-0.2cm}\psfig{file=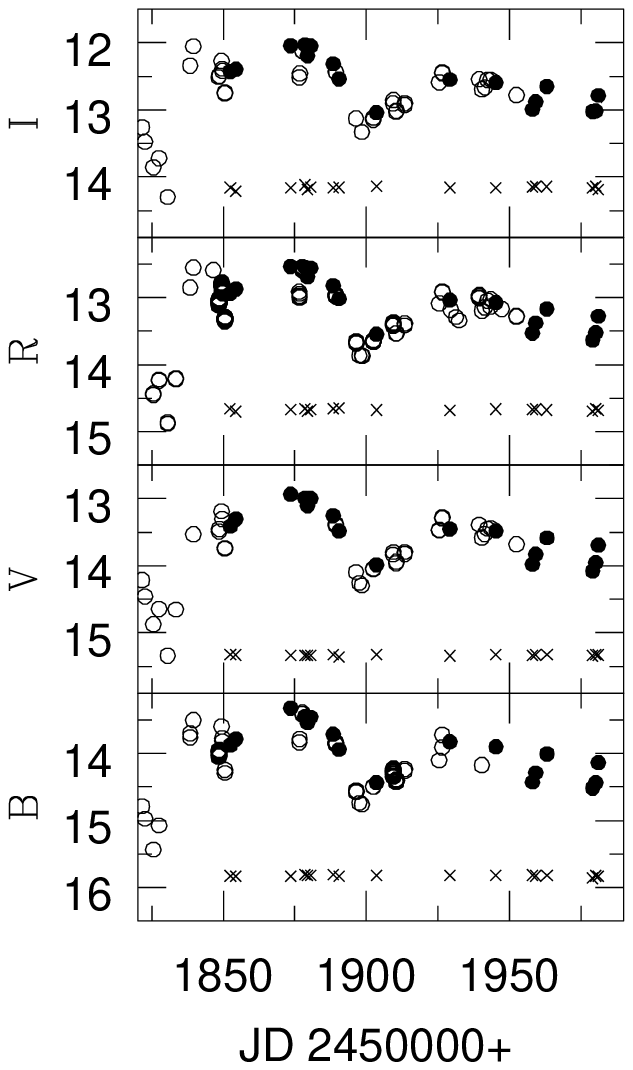} 
\caption{Filled circles show the long-term lightcurves of 
S5 0716$+$714 in the BVRI bands from our
observations between November 2000 and March 2001 (Sect.\ 3.1.4). 
The crosses refer to our measurements of the standard star 6 defined
by Villata et al.\ (1998) plotted after adding offsets of 1.6, 1.7, 1.4 and 
1.2 magnitudes in the B,V,R and I filters, respectively.  Data taken from 
the long-term monitoring observations by Raiteri et al.\ (2003)
are shown as open circles.
}
\end{figure}

\subsection{S5 0716$+$714}
The results of the INOV campaign are given in Table 2
where we list the observing date and the filter(s) used, the number of
data points in the DLC, the mean and standard deviation of the blazar DLC, the standard
deviation of the comparison stars' DLCs, the variability classification 
for the blazar and the confidence level of any observed variability. 
The object shows clear evidence for 
INOV on two nights,  7 and 14 April, 1996. The DLCs are shown in 
Fig.\ 1.

For some of the densely monitored blazars, the substantial 
individual optical flares seen on time scale of $\sim$1 day are
characterised by exponential profiles for rise and  decay in flux. 
Examples include
the intraday variable blazars PKS 2155$-$304 (Urry et al.\ 1993) and
S5 0954$+$658 (Wagner et al.\ 1993). For S5 0716$+$716, too, the individual flares 
on day-like time scales recorded in the 1995$-$96 optical monitoring  
are well described by an exponential flux change (Sagar et al.\ 1999). 

During our observations (Table 2), S5 0716$+$714 showed  INOV with
rates $\ge$ 0.01 mag/hr on two nights. Since the variability on both nights 
indicate roughly linear trends on the magnitude scale, we have fitted least-squares 
regression lines to the variable segment of the DLCs. These best fit
lines are shown in Fig.\ 2 and their slopes are given in Table 3 together with
the regression coefficients ($\gamma$). The rate of brightening 
on 7 April 1996 is 
faster than the fading rate on 14 April 1996, but  
 on each night, the slopes are essentially the same in both 
V and {\it R} bands. 
Thus, at least on the two nights, 
the prominent events of INOV can be described by linear
trends on the magnitude scale, which therefore correspond to 
exponential flux variations. 

\hspace*{-0.8cm}{
\begin{table}
\caption{Results of the intra-night monitoring of S5 0716$+$714 and BL Lacertae
 (see Sect. 3)}
\begin{tabular}{lcrcclr} \hline
\noalign{\smallskip}
Date    & Band     & N & Diff. mag & $\sigma$ (DLC) & Var? & C \\
        &          &   & BL$-$S1 & S1$-$S2     &      &   \\
        &          &   &   mag     &      mag       &      &   \\
\noalign{\smallskip}
\hline
\noalign{\smallskip}
                  & &   &                  &       &    &     \\
{\bf 0716$+$714} & &   &                  &       &    &    \\
                  & &   &                  &       &    &     \\
04 Apr.96 & V & 14 & 0.88 $\pm$ 0.01  & 0.008 & NV &      \\
05 Apr.96 & R & 35 & 0.75 $\pm$ 0.02  & 0.008  & NV &     \\
07 Apr.96 & V & 12 & 0.73 $\pm$ 0.02  & 0.007  & V  & 2.9 \\
          & R & 12 & 0.68 $\pm$ 0.02  & 0.007  & V  & 2.9 \\
10 Apr.96 & V & 13 & 0.82 $\pm$ 0.01  & 0.008  & NV &     \\
          & R & 13 & 0.82 $\pm$ 0.01  & 0.009  & NV &     \\
11 Apr.96 & V & 16 & 0.91 $\pm$ 0.01  & 0.008  & NV &     \\
          & R & 17 & 0.90 $\pm$ 0.01  & 0.012  & NV &     \\
12 Apr.96 & V & 09 & 0.75 $\pm$ 0.01  & 0.007  & NV &     \\
          & R & 17 & 0.73 $\pm$ 0.01  & 0.005  & NV &     \\
13 Apr.96 & V & 10 & 0.62 $\pm$ 0.01  & 0.007  & NV &     \\
          & R & 10 & 0.60 $\pm$ 0.01  & 0.004  & NV &     \\
14 Apr.96 & V & 18 & 0.52 $\pm$ 0.02  & 0.006  & V  & 3.3 \\
          & R & 18 & 0.50 $\pm$ 0.03  & 0.011  & V  & 2.7 \\
18 Apr.96 & V & 16 & 0.64 $\pm$ 0.02  & 0.012  & NV &     \\
          & R & 15 & 0.60 $\pm$ 0.01  & 0.009  & NV &     \\
19 Apr.96 & V & 12 & 0.81 $\pm$ 0.02  & 0.009  & NV &     \\
          & R & 12 & 0.76 $\pm$ 0.02  & 0.008  & NV &     \\
20 Apr.96 & V & 10 & 0.93 $\pm$ 0.01  & 0.006  & NV &     \\
          & R & 09 & 0.88 $\pm$ 0.01  & 0.006  & NV &     \\ \hline
	  &   &    &                  &        &    &   \\
{\bf BL Lac} & &   &                  &        &    &     \\
                  & &   &                  &       &    &     \\
19 Oct.01 & B & 10 & 0.56 $\pm$ 0.04  & 0.004  & V  & 10.0    \\
          & V & 10 & 0.55 $\pm$ 0.02  & 0.009  & V  & 2.2    \\
          & R & 10 & 0.37 $\pm$ 0.04  & 0.004  & V  & 10.0    \\
20 Oct.01 & B &  8 & 0.60 $\pm$ 0.04  & 0.008  & V  & 5.0 \\
          & V &  8 & 0.60 $\pm$ 0.01  & 0.001  & V  & 10.0 \\
          & R &  8 & 0.42 $\pm$ 0.02  & 0.009  & V  & 2.2 \\
21 Oct.01 & B & 10 & 0.20 $\pm$ 0.01  & 0.008  & NV &  \\
          & V & 10 & 0.21 $\pm$ 0.01  & 0.004  & NV &  \\
          & R & 10 & 0.04 $\pm$ 0.01  & 0.015  & NV &  \\
22 Oct.01 & B & 10 & 0.43 $\pm$ 0.04  & 0.003  & V  & 7.5 \\
          & V & 10 & 0.43 $\pm$ 0.04  & 0.002  & V  & 6.7 \\
          & R & 10 & 0.24 $\pm$ 0.04  & 0.013  & V  & 6.0 \\
25 Oct.01 & B &  8 & 0.73 $\pm$ 0.005 & 0.003  & NV & 2.5 \\
          & V &  7 & 0.70 $\pm$ 0.002 & 0.001  & NV & 2.0 \\
          & R &  7 & 0.50 $\pm$ 0.004 & 0.002  & NV  & 5.0 \\

\noalign{\smallskip}
\hline
\end{tabular}
\end{table}
}

\subsubsection{Determination of the time lag}
To check for any time lag between {\it V} and {\it R} band DLCs we
have used both the Interpolated Cross-Correlation Function (ICCF) 
method (Gaskell \& Peterson  1987)
and the Discrete Correlation function (DCF) method (Edelson \& Krolik 1988).
In the DCF we first calculated the set of {\it unbinned discrete correlation functions}
defined as

\begin{equation}
UDCF_{ij} = \frac{(a_i - a)(b_j - b)}{\sigma_a*\sigma_b},
\end{equation}

\noindent where $a_i$ and $b_i$ are the observed differential magnitudes 
in {\it V} and {\it R} bands, and  $a, b, \sigma_a$ and $\sigma_b$ are, respectively, the means and standard deviations of the DLCs in the {\it V} and
R bands. Binning the result in $\tau$ gives the DCF($\tau$). Averaging
over M pairs for which $\tau - \delta \tau /2 \le \delta t_{ij} < \tau + \delta \tau/2$
gives

\begin{equation}
DCF(\tau) = UDCF_{ij}/M.
\end{equation}

The errors in the DCF are calculated using
\begin{equation}
\sigma_{DCF}(\tau) = \frac{1}{(M-1)}{\Sigma[UDCF_{ij} - DCF(\tau)]}^{1/2}.
\end{equation}

\begin{table}
\caption{Slopes and regression coefficients ($\gamma$) of the linear 
least-squares fit to the intra-night magnitude variations of S5 0716$+$714}
\begin{tabular}{cccccc} \hline
\noalign{\smallskip}
Date &  Band   &  DLC               & $\gamma$ & DLC                &  $\gamma$  \\
     &         &  BL$-$S1           &          & BL$-$S2            &            \\
     &         & slope $\pm \sigma$ &          & slope $\pm \sigma$ &            \\      
\noalign{\smallskip}
\hline
\noalign{\smallskip}
07 Apr.96  &  V   & -0.033 $\pm$ 0.012  &  0.65  &  -0.052 $\pm$ 0.011  &  0.84  \\
           &  R   & -0.033 $\pm$ 0.012  &  0.68  &  -0.039 $\pm$ 0.008  &  0.84  \\
14 Apr.96  &  V   &  0.024 $\pm$ 0.003  &  0.88  &   0.026 $\pm$ 0.003  &  0.94  \\
           &  R   &  0.023 $\pm$ 0.004  &  0.83  &   0.036 $\pm$ 0.004  &  0.89  \\
\noalign{\smallskip}
\hline
\end{tabular}
\end{table}

The computed DCF and ICCF for the two nights when the object showed 
INOV are displayed in Fig.\ 3.  On both these nights there are indications
that the correlation functions peak at small time lags (about
6 and 13 minutes, respectively, for 7 and 14 April 1996) with the variation 
at V leading that at R on both dates.  But since the sampling is relatively sparse,
with the measurement intervals close to these putative lags, they
might not be significant and hence should be treated
with caution. A similar lag of $\sim$6 min between V and I bands was
also reported by Qian et al.\ (2000) from observations on a single night, but their
data may be best interpreted as providing an upper limit to the lag of that magnitude.
Note that Villata et al.\ (2000) present a strict
upper limit of $\sim$10 min to a possible delay between B and I band
variations using high quality data, densely sampled on a single night.

\subsubsection{Colour variations}
The instrumental magnitudes in {\it V} and {\it R} are converted to the standard
system by carrying out differential photometry using star 5 of 
Villata et al.\ (1998).
To see if the source exhibited any colour variations on inter-night 
time scales we then computed V$-$R 
colours by interpolating the {\it R} band magnitudes to the times of the {\it V} exposures
and computed mean colour, $\langle$V$-$R$\rangle$, and 
the mean magnitude, $\langle$V$\rangle$, 
for each night. The colour--magnitude 
diagram (CMD) thus obtained, is shown in Fig.\ 4.  
The solid line in 
the figure shows the unweighted linear least-squares fit to  the data 
points, which has 
a slope of 0.01 $\pm$0.04 and a correlation 
coefficient of 0.15. Thus, no significant evidence for  
colour variation with brightness is evident. This is also true for the
two nights when INOV was observed. 
 Similar colour analyses on time scales comparable to those of our 
observations of the optical light curves from 1994$-$2001 of this source by 
Raiteri et al.\ (2003) show at most  weak correlations between colour index and source brightness, in the sense that
the spectrum sometimes steepens as the brightness decreases. 

\subsubsection{Inter-night variability}
When compared with the available long-term light curve (Raiteri et al. 2003; 
Nesci et al. 2005), the blazar 
was  in relatively quiescent state during our 2 weeks of
monitoring. Still, the observed variations are often
 much larger than the rms scatter of the relative magnitudes of the pair of
comparison stars, which is found to be $\sim$ 0.008 mag for both passbands. 
There
were two prominent variations (around JD 2450180 and JD 2450188 with 
amplitudes of about 0.2 and 0.4 mag, respectively) recorded during the present 
observations and shown in Fig.\ 5, in which we plot two data points 
per night, one from the start and the other from the end of the observation.
 By comparing our data (filled circles) with those taken from Raiteri 
et al. (2003; open circles), it is clear that the two datasets 
are generally consistent and complementary to each other.  
The combined dataset reveals 
much finer scale structures in the lightcurves than 
is evident from either dataset.

\begin{figure*}
\centering
\includegraphics[width=16cm]{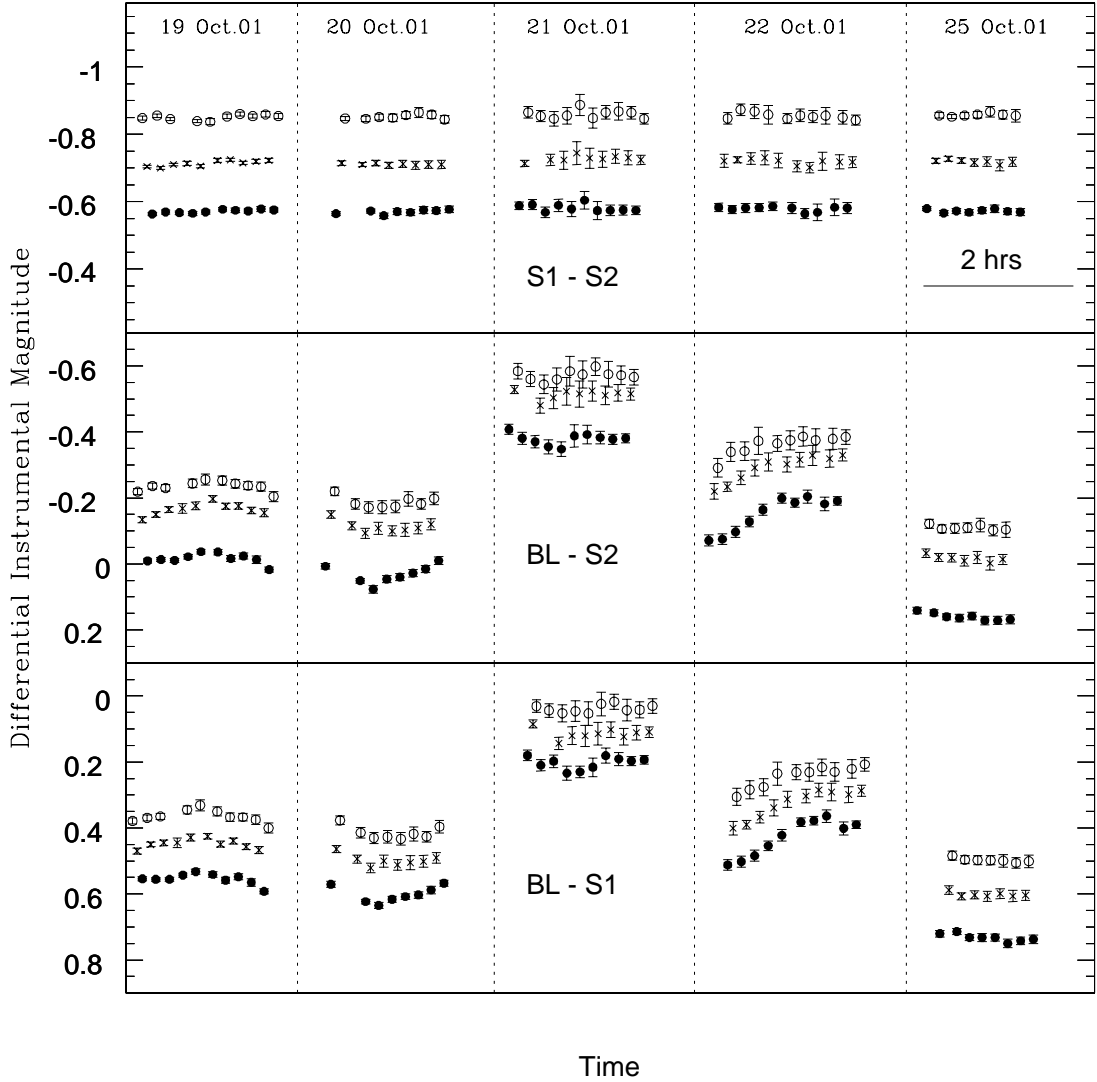} 
\caption{DLCs for BL Lacertae and comparison stars in the bands 
{\it B} (filled circles), {\it V} (crosses) and {\it R} (open circles). The {\it V} and {\it R} band 
differential magnitudes are shifted by constant amounts for 
clarity of display. The time scale of all the DLCs is marked with 
the horizontal bar of 2 hours duration.}
\end{figure*}

\subsubsection{Long-term variability}

We made further observations of this blazar on 
17 nights between November 2000 and March 2001 (one frame each in BVRI filters 
on each night) to examine its long-term variability. 
The long-term lightcurve is shown in Fig.\ 6. 
An important feature in this light curve is a major flare seen
around JD 2451875 in all bands. 
These variations were achromatic to within the errors as a regression analysis
gives a linear correlation coefficient of only 0.35.  In Fig.\ 6 we
have also plotted with open circles, the measurements reported
by Raiteri et al.\ (2003). Combination of our and their data better
defines the shape of the flare. Also, it clearly shows
a second shallower bump following the flare mentioned above. 
Note that the variations shown in Fig.\ 6 coincide with the most
active phase seen in the long-term lightcurve covering the period
1994$-$2001 (Raiteri et al.\ 2003).  A sample of our
our observations of S5 0716+714 are shown in Table 4; the complete data are
available as
an electronic table.

\begin{table}
\caption{Results of photometric monitoring observations
of S5 0716+714. The magnitudes are not corrected for
galactic extinction. The table in full is available electronically.}
\begin{tabular}{cccc} \hline
\noalign{\smallskip}
Julian Date  &  Filter   &     m          &  $\sigma_m$         \\
             &           &  (mag)    &  (mag)               \\
\noalign{\smallskip} \hline
\noalign{\smallskip}
2451631.1532 & B & 15.01  & 0.01  \\
2451631.1625 & V & 14.51  & 0.02  \\
2451631.1667 & R & 14.02  & 0.01  \\
2451631.1700 & I & 13.23  & 0.05  \\
2451632.1100 & R & 13.92  & 0.01  \\
2451632.1166 & B & 14.86  & 0.01  \\
2451632.1248 & V & 14.38  & 0.02  \\
2451632.1292 & I & 13.35  & 0.05  \\
2451633.0966 & R & 13.90  & 0.01  \\
2451633.0994 & I & 13.33  & 0.05  \\
\noalign{\smallskip}
\hline
\end{tabular}
\end{table}

\begin{table}
\caption{Results of photometric monitoring observations
of BL Lacertae. The magnitudes are not corrected for galactic
extinction. The table in full is available electronically.}
\begin{tabular}{cccc} \hline
\noalign{\smallskip}
Julian Date  &  Filter   &     m          &  $\sigma_m$         \\
             &           &  (mag)    &  (mag)               \\
\noalign{\smallskip} \hline
\noalign{\smallskip}
 2452202.0390 & R  & 14.07   & 0.03  \\
 2452202.0418 & V  & 14.76   & 0.03  \\
 2452202.0449 & B  & 15.64   & 0.03  \\
 2452202.0478 & R  & 14.06   & 0.03  \\
 2452202.0499 & V  & 14.74   & 0.03  \\
 2452202.0529 & B  & 15.65   & 0.03  \\
 2452202.0557 & R  & 14.06   & 0.03  \\
 2452202.0577 & V  & 14.74   & 0.03  \\
 2452202.0611 & B  & 15.65   & 0.03  \\
 2452202.0658 & V  & 14.74   & 0.03  \\
\noalign{\smallskip}
\hline
\end{tabular}
\end{table}

\subsection{BL Lacertae}

This blazar showed INOV in at least two bands on three of the five
nights it was monitored by us. The summary of the results   
is given in Table 2, the DLCs are shown in Fig.\ 7 and the 
complete data are available electronically (Table 5).
Our observations overlap with the long-term photometric
measurements reported by Villata et al. (2004) and the two
datasets are in good agreement.
The first night we observed this source, 19 October 2001, it
was found to decline in brightness toward the end of the observations. On the
following night, the source increased in brightness by
$\sim$ 0.07 {\it B} mag toward the end of the observations. When BL Lac was observed
on 21 October, it was found to be brighter by $\sim$ 0.4 {\it B} mag than on the
previous night. On the following night the source was found to be
about 0.3 mag fainter than at the end of the previous night, but it increased in
brightness by $\sim$ 0.15 {\it B} mag within $\sim$ 1.4 hr, with the other bands
rising by smaller amounts. On the last night of
our observations, 25  October, the source had faded by $\sim$ 0.3 mag
compared to its brightness recorded three days earlier.  

\subsubsection{Colour variations}
The object
showed hints for colour variations (colours are estimated
as described in Sect.\ 3.1.2, using stars B and C of Smith et al.\ 1985) on 
inter-night time scales, 
as seen from the CMD in Fig.\ 8. 
The solid line is the unweighted
linear least-squares fit to the data, giving a slope
of 0.05 $\pm$ 0.03 and a linear correlation coefficient of 0.67. 
Thus, our observations on inter-night time scales suggest a possible bluer
when brighter trend for BL Lacertae. However, removing the first observed point
at V $\sim$14.9 mag, the unweighted
linear least-squares fit gives a slope of 0.01 and a linear correlation
coefficient of 0.39, which is compatible with a constant V$-$R colour index with
brightness.
Fig.\ 9 shows the CMD on the night of 22 October 2001, when
the object showed clear evidence of INOV. Linear least-squares fit to the 
data gives
a slope of 0.26 $\pm$ 0.04 with a linear correlation coefficient
of 0.91. Thus, on intra-night time scales, the object showed a
clear bluer when brighter correlation.  Recently, Villata et al. (2004) have
noticed that BL Lacertae shows a stronger CMD on intra-day
time scales, compared to longer (month-like) time scales. 

\subsubsection{Determination of time lag}
A cross correlation function analysis (see Sect. 3.1.1) was performed to 
look for any
time lag between the {\it V} and {\it R} band variations on the one night (22 October 2001) when 
the object showed INOV in all the three bands. The {\it V} band was found to lead
the {\it R} band variations by about 3 minutes based on both ICCF and DCF 
analysis. However, given the noise on individual points, this time lag
is not significant (Fig.\ 10). 

\section{Summary}
Based on the  published long-term light curves of both these blazars 
(Raiteri et al.  2003; Villata et al.\ 2004; Nesci et al.\ 2005), the 
present INOV observations 
are seen to coincide with 
their relatively faint optical states. Nonetheless, two  
clear events 
of INOV, one of brightening, and the other of fading,
were observed in S5 0716$+$714. The events are well fitted with
an exponential intensity profile with essentially the same rate of 
variation in the {\it V} and {\it R} bands. 
We note a hint of a possible  temporal lag of $\sim$10 minutes between these 
two bands 
with the shorter wavelength leading the longer, as
expected in the standard shock-in-jet model; 
such a lag is not
expected in most disk-based variability models (e.g., Wiita 2005). 
In the present eight nights of our monitoring of S5 0716$+$714 spanning  
about a fortnight, two noticeable events of inter-night variability (one 
around JD 2450180 
and the other around JD 2450188) of amplitude $\ge$ 0.2 mag were
found. Our long-term 
monitoring of this blazar, on the other hand, coincided with the maximum 
brightness
phase seen in the lightcurve covering 1994$-$2001, as reported
by Raiteri et al. (2003). By combing their data with the 
present measurements a large 1.5 magnitude flare is detected
around JD 2451875 lasting for about 75 days. This 
is followed by a weaker variation on a similar timescale.
No clear evidence of colour variation with brightness was found in either
our inter-night or our intra-night monitoring of S5 0716+714.

The other blazar, BL Lacertae, was found to vary strongly in all 
three bands (B,V,R), 
on one of the five nights we monitored it and in two of the three bands on
two other nights. Also, on the night when strong INOV was observed in all three
bands, a temporal lag between the {\it V} and {\it R} band lightcurves may be present,
with the {\it V} band variations apparently leading the {\it R} band variations by a few minutes.
Nesci et al. (1998) also noted that no time lag between different
bands could be unambiguously detected during intra-night monitoring observations
of BL Lac in 1997.  A clear trend was seen in our data for the source to 
become bluer when brighter on intra-night time scales. Such a
clear trend was not evident on inter-night time scales, although the data suggest one
may be present.
The presence (or absence) of this consistency between the colour-magnitude
behaviour  on intra-night and inter-night time scales can provide 
interesting clues about the origin of blazar variability from
hour-like to much longer time sales.  
In view of this, sustained efforts
for multi-colour monitoring of blazars on different time scales will be 
fruitful.

\begin{figure}
\centering
\includegraphics{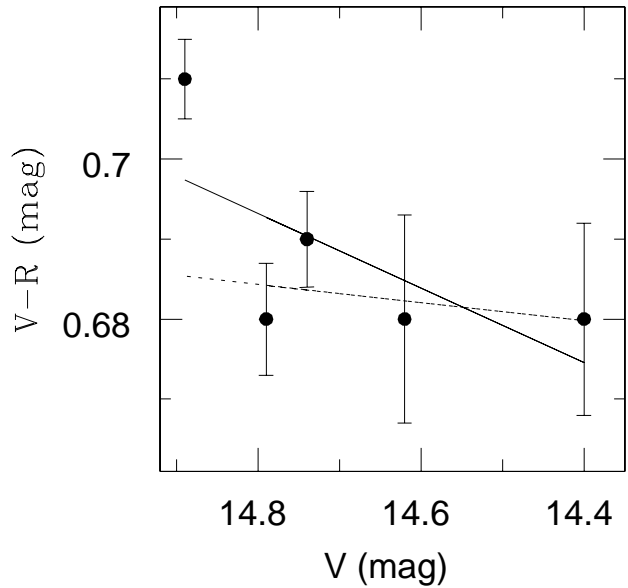} 
\caption{Inter-night CMD for BL Lacertae based on our observations in 2001. 
The solid line is the linear least-squares fit to the data points. The dotted
line is the linear least-squares fit to the data excluding 
the first data point at V $\sim$14.9 mag.}
\label{Colour}
\end{figure}

\begin{figure}
\centering
\includegraphics{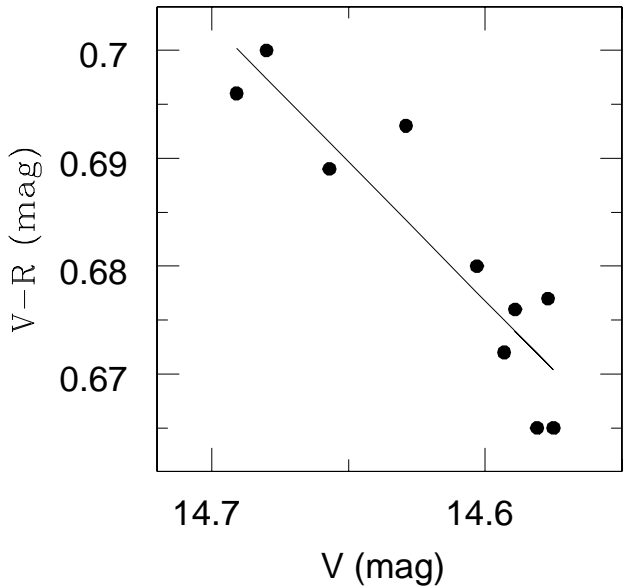} 
\caption{Intra-night CMD for BL Lacertae on 22 October 2001. The solid
line is the linear least-squares fit to the data points}
\end{figure}

\begin{figure}
\hspace*{-0.9cm}\psfig{file=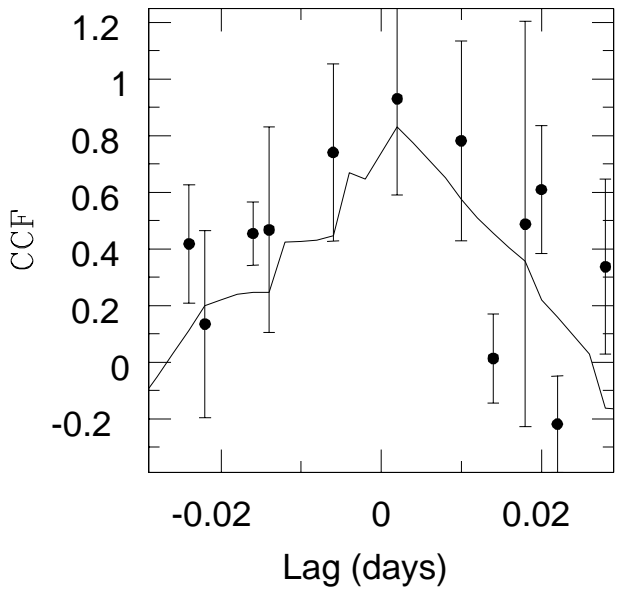}
\caption{Cross correlation function for the {\it V} and {\it R} band lightcurves of BL Lacertae on
22 October 2001. Full circles with error bars refer to the DCF, whereas the solid line
is calculated from the ICCF (see Sect.\ 3.1.1).}
\end{figure}

\section*{Acknowledgments}
We thank the anonymous referee for suggestions which significantly improved
the presentation of our results. 
This research has made use of the NASA/IPAC Extragalactic Database (NED)
which is operated by the Jet Propulsion Laboratory, under contract with
the National Aeronautics and Space Administration. 
PJW acknowledges support from NSF grant AST-0507529 and continuing hospitality
at the Department of Astrophysical Sciences, Princeton University.

\end{document}